\documentclass[]{spie}  

\usepackage{amsmath,amsfonts,amssymb}
\usepackage{graphicx}
\usepackage[colorlinks=true, allcolors=blue]{hyperref}
\usepackage{amsmath,amsfonts,amssymb}
\usepackage{multicol}
\usepackage{tabu}
\usepackage{float}
\usepackage{textcomp,gensymb}

\title{Characterization of a high efficiency silicon photomultiplier for millisecond to sub-microsecond  astrophysical transient searches}

\author[a]{Siyang Li}
\author[a-h]{George F. Smoot}

\affil[a]{Department of Physics, University of California, Berkeley, USA}
\affil[b]{Lawrence Berkeley National Laboratory, USA, {\it Emeritus}}
\affil[c]{Department of Physics, Hong Kong University of Science and Technology, China}
\affil[d]{Institute for Advanced Study, Hong Kong University of Science and Technology, China}
\affil[e]{Energetic Cosmos Laboratory, Nazarbayev University, Kazakhstan}
\affil[f]{Department of Physics, Universit\'e Paris Diderot, France {\it Emeritus}}
\affil[g]{Paris Centre for Cosmological Physics, Universit\'e Paris, France}   
\affil[h]{Donostia International Physics Center, Universidad del Pa\'is Vasco, Spain}

\authorinfo{Further author information: (Send correspondence to S.L.)\\S.L.: E-mail: seanli@berkeley.edu}

\begin{document} 
\maketitle

\begin{abstract}

We characterized the S14160-3050HS Multi-Pixel Photon Counter (MPPC), a high efficiency, single channel silicon photomultiplier manufactured by Hamamatsu Photonics K.K. All measurements were performed at a room temperature of (23.0 $\pm$ 0.3) \textdegree C. We obtained an I-V curve and used relative derivatives to find a breakdown voltage of 38.88 V. At a 3 V over voltage, we find a dark count rate of 1.08  MHz, crosstalk probability of 21 $\%$, photon detection efficiency of 55 $\%$ at 450 nm, and saturation at 1.0x10\textsuperscript{11} photons per second. The S14160-3050HS MPPC is a candidate detector for the Ultra-Fast Astronomy (UFA) telescope which will characterize the optical (320 nm - 650 nm) sky in the millisecond to sub-microsecond timescales using two photon counting arrays operated in coincidence on the 0.7 meter Nazarbayev University Transient Telescope at the Assy-Turgen Astrophysical Observatory (NUTTelA-TAO) located near Almaty, Kazakhstan. We discuss advantages and disadvantages of using the S14160-3050HS MPPC for the UFA telescope and future ground-based telescopes in sub-second time domain astrophysics.
\end{abstract}

\keywords{silicon photomultiplier, instrumentation, telescope, detector characterization, high efficiency, astrophysical transients, quantum optics}

\section{INTRODUCTION}

Silicon photomultipliers (SiPM) are solid-state photodetectors consisting of thousands of single-photon avalanche photodiodes (SPAD), or microcells, connected in parallel. A photon incident on a microcell triggers a Geiger avalanche breakdown which produces an output current. The output of a SiPM is the sum of outputs for each individual microcell.


SPAD and SiPM are currently being implemented by several groups to investigate the variability of a variety of astrophysical phenomena. The Nakamori et al. (2020)\cite{Nakamori_2020}, Optical Pulsar Timing Analyser (OPTIMA)\cite{Straubmeier:2001eb}, Silicon Fast Astronomical Photometer (SiFAP)\cite{Ambrosino_2017trt},
and Aqueye+ \cite{Zampieri_2019_10.1093/mnrasl/slz043} teams have used SiPM to observe millisecond pulsars, and the Extreme Universe Space Observatory onboard the Japanese Experiment Module (JEM-EUSO) will characterize atmospheric events such as high energy cosmic ray showers using a SiPM array (Mini-EUSO) aboard the International Space Station\cite{Capel_2018AdSpR..62.2954C}. The Cherenkov Telescope Array (CTA) will use ground-based SiPM telescopes to search for high energy gamma rays\cite{Ambrosino_2017trt}, and the Pulsed All-sky Near-infrared Optical Search for Extreterrestrial Intelligence (PANOSETI) collaboration is currently developing a SiPM all-sky optical and SPAD wide-field near-infrared observatory to search for extraterrestrial technosignatures \cite{Wright_2018_10.1117/12.2314268, Li_2019_NIRDAPD_10.1117/12.2519207}.


The first-generation Ultra-Fast Astronomy (UFA) observatory\cite{Li_2019a} will characterize the optical (320 nm - 650 nm) sky in the millisecond to sub-microsecond timescales using two photon counting arrays operated in coincidence on the 0.7 meter Nazarbayev University Transient Telescope at the Assy-Turgen Astrophysical Observatory (NUTTelA-TAO) located near Almaty, Kazakhstan. We have narrowed down our candidate analog SiPM to three Multi-Pixel Photon Counters (MPPC) from Hamamatsu Photonics K.K.: S13360-3050CS MPPC, S14520-3050VS MPPC, and S14160-3050HS MPPC. We are also monitoring other sources including position sensitive SiPMs.

Our goal in this work was to both refine our testing methods to prepare for further developments and characterize the S14160-3050HS MPPC using the same methods used to characterize the S13360-3050CS MPPC for the UFA telescope in a previous study\cite{Li_2019a} to provide an equivalent baseline comparison.  We present a characterization of the breakdown voltage, dark count rate, crosstalk probability, photon detection efficiency, and saturation of the S14160-3050HS MPPC and discuss advantages and disadvantages of using this SiPM for the UFA telescope and future ground-based telescopes in sub-second time domain astrophysics.


\section{EXPERIMENTAL SETUP}

\begin{figure} [ht]
\begin{center}
\begin{multicols}{2}
\includegraphics[height=6cm]{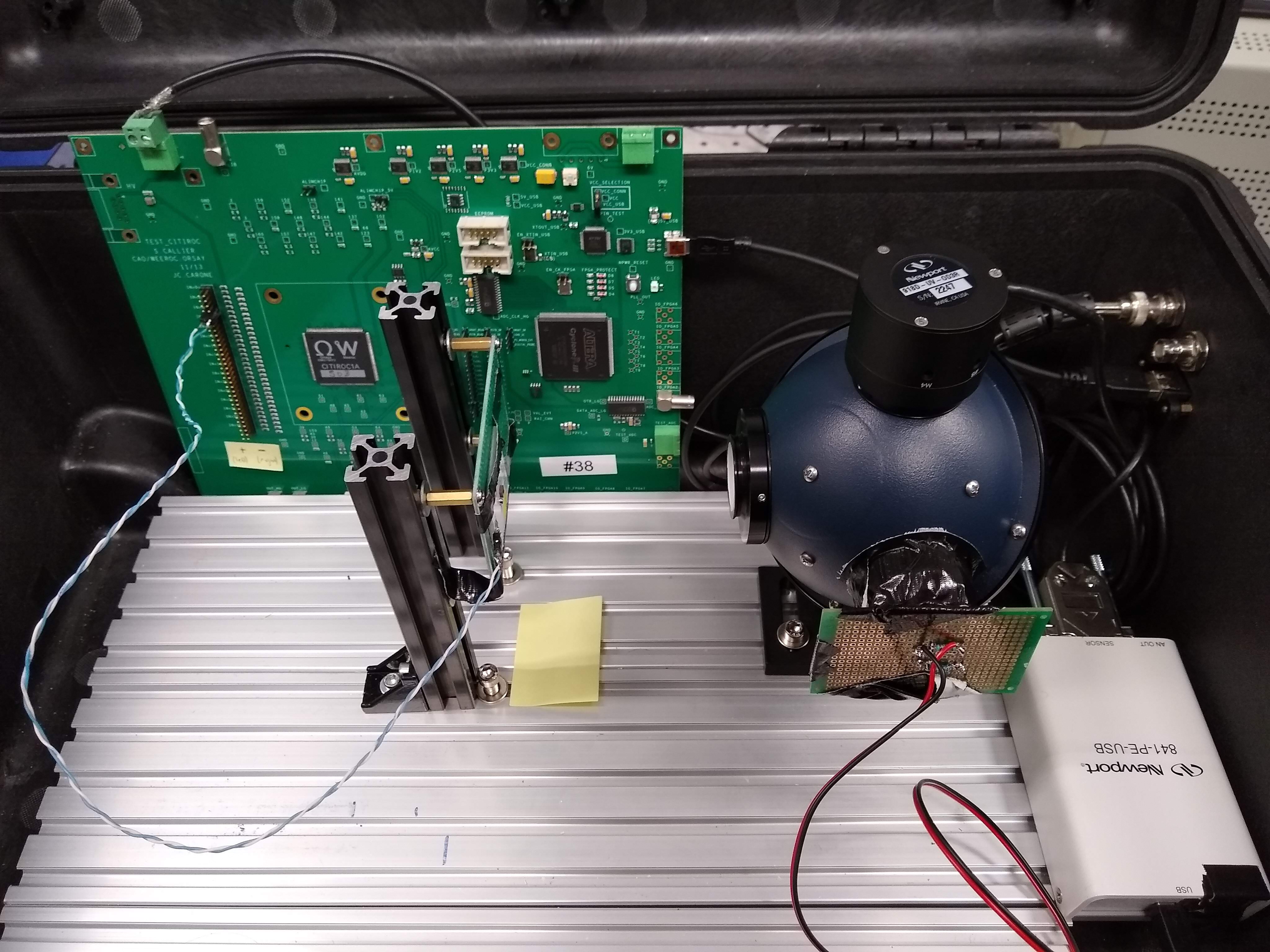}
\includegraphics[height=6cm]{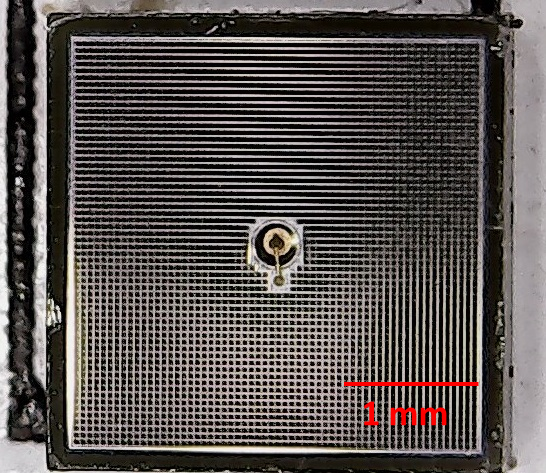}\par 
\end{multicols}
\end{center}
\caption{LEFT: Dark box test stand used to characterize the S14160-3050HS MPPC in this study. RIGHT: S14160-3050HS MPPC from Hamamatsu.}
\label{fig:darkbox}
\end{figure}

The experimental setup and S14160-3050HS MPPC characterized in this study can be seen in Fig. \ref{fig:darkbox}. 
Detector parameters provided by Hamamatsu\cite{datasheet} can be found in Table \ref{tab:parameters}. The SiPM was placed inside a light-tight dark box containing BNC, USB, and power supply bulkhead feedthroughs. A 3.3 inch integrating sphere (Newport 819D-SL-3.3) was used to uniformly scatter light emitted from 365 nm, 400 nm, 450 nm, 525 nm, 560 nm, 660 nm, 720 nm, 810 nm, and 900 nm light-emitting diodes (LEDs) when illuminating the SiPM. The SiPM was placed 100 mm from the exit port of the integrating sphere. An internal baffle between the two side ports inside the integrating sphere prevents light emitted from the LEDs from directly exiting the integrating sphere before scattering. The LEDs were pulsed at 900 kHz using a waveform generator (Tektronix AFG1062) for photon detection efficiency (PDE) measurements and biased with direct current for linearity and saturation measurements.

We mounted a NIST calibrated ultraviolet enhanced silicon photodiode (Newport 918D-UV-OD3R) on the northern port of the integrating sphere to monitor for changes in irradiance during measurements. A second photodiode of the same model was placed at the location of the SiPM to determine the ratio of irradiances between the northern port of the integrating sphere and the location of the SiPM. This ratio was recalibrated for each LED. We used the 32-channel Citiroc 1A evaluation board manufactured by Weeroc for our readout system. Current was measured using a picoammeter (Tektronix PWS2721). All measurements were performed at a room temperature of (23.0 $\pm$ 0.3) \textdegree C.

\begin{table}[H]
\begin{center}
\begin{tabu}  { | X[c] | X[c] | }

\hline
\textbf{Parameter} &
\textbf{S14160-3050HS MPPC} \\
\hline
Spectral Range & 270 nm - 900 nm \\
\hline
Photosensitive Area  & 3.0 mm x 3.0 mm \\
\hline
Number of Microcells & 3531 \\
\hline
Pixel Pitch  & 50 $\mu$m \\
\hline
Fill Factor & 74 $\%$ \\
\hline
\end{tabu}
\end{center}
\caption{Detector parameters for the S14160-3050HS MPPC characterized in this study.}
\label{tab:parameters}
\end{table} 


\section{Results}

\subsection{Breakdown Voltage}

The breakdown voltage of a SiPM is the minimum bias voltage needed to produce an electric field in the depletion region sufficient to sustain avalanche breakdowns and is qualitatively the threshold above which the dark current of the SiPM significantly increases. SiPMs are typically operated in Geiger mode at an over voltage defined as the positive difference between the operating bias and breakdown voltage. Detector characteristics such as dark count rate, crosstalk probability, and PDE are functions of over voltage. 

\begin{figure} [H]
\begin{center}
\begin{tabular}{c} 
\includegraphics[height=6cm]{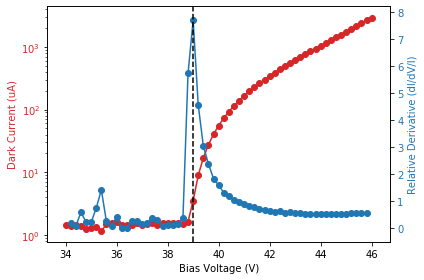}
\end{tabular}
\end{center}
\caption[example] 
{Dark current (red) and its relative derivative (blue) as functions of bias voltage. The vertical black dashed line marks the location of the maximum relative derivative. Data points shown are in 0.2 V increments.}
\label{fig:Breakdown}
\end{figure}

To find the breakdown voltage for this particular S14160-3050HS MPPC, we first measured its dark current as a function of bias voltage in 0.2 V increments from 34.00 V to 46.00 V. We calculated the relative derivative  $dI/dV/I$ at each point excluding 34.00 V and located its maximum to obtain the breakdown voltage to 0.2 V precision. We again measured the dark current as a function of bias voltage, this time in in 0.01 V increments and a range of $\pm$0.2 V around the maximum relative derivative found before, and located the new maximum relative derivative to obtain the breakdown voltage to 0.01 V precision. An image of the relative derivative superimposed on a logorithmic I-V plot can be seen in Fig. \ref{fig:Breakdown}. We find a breakdown voltage of 38.88 V.


\subsection{Dark Count Rate}

\begin{figure}[H]
\centering
\includegraphics[height=6cm]{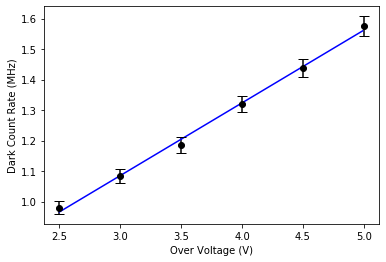}
\caption{Mean dark count rate as a function of over voltage. Data points are shown in black and a linear least squares fit is shown by the blue line.}
\label{fig:DarkCount}
\end{figure}

SiPM dark counts are primarily caused by thermally generated photoelectrons that trigger avalanche breakdowns. Because dark counts are indistinguishable from photon events, a high dark count rate can limit the resolution of a SiPM telescope. Using a 0.5 photoelectron threshold, we measured a mean dark count rate of (1.08 $\pm$ 0.03) MHz at a 3 V over voltage. Using a linear least squares fit, we find a slope of (238 $\pm$ 7) kHz/V. The mean dark count rate as a function of over voltage can be seen in Fig. \ref{fig:DarkCount}.

\subsection{Crosstalk Probability}

\begin{figure} [H]
\begin{center}
\begin{tabular}{c} 
\includegraphics[height=6cm]{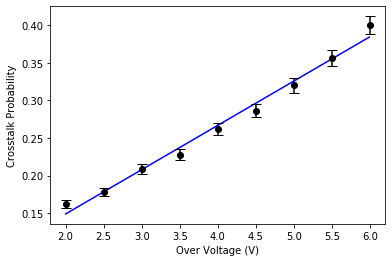}
\end{tabular}
\end{center}
\caption[example] 
{Crosstalk probability as a function of over voltage. Measurements are shown in black and a linear least squares fit is shown by the blue line.}
\label{fig:Crosstalk}
\end{figure}

Crosstalk occurs when the acceleration of photoelectrons inside a microcell generates photons that trigger avalanche breakdowns in nearby microcells. We assume the number of multi-microcell crosstalk events and two photoelectron and above dark count events are negligible and measured the crosstalk probability in the absence of light by dividing the 2 photoelectron and higher count rate by the 1 photoelectron and higher count rate obtained using 1.5 and 0.5 photoelectron thresholds, respectively. We measured a crosstalk probability of (21 $\pm$ 3) $\%$ at a 3 V over voltage. Using a linear least squares fit, we find a slope of (5.9 $\pm$ 0.2) $\%$/V. The crosstalk probability as a function of over voltage can be seen in Fig. \ref{fig:Crosstalk}.  We note that the method used here does not account for the effects of afterpulsing and will overestimate the number of independent crosstalk events. 

\subsection{Photon Detection Efficiency}

The PDE is defined as the ratio between the number of detected photons and the number of incident photons and can be expressed as

\begin{equation}
    PDE = f * \eta * P_g
\end{equation}

where $f$ is the geometrical fill factor, $\eta$ is the quantum efficiency, and $P_g$ is the Geiger efficiency.

\begin{figure} [H]
\begin{center}
\begin{multicols}{2}
\includegraphics[height=5.5cm]{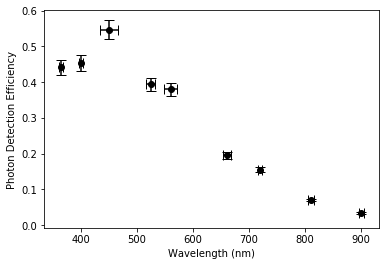}
\includegraphics[height=5.5cm]{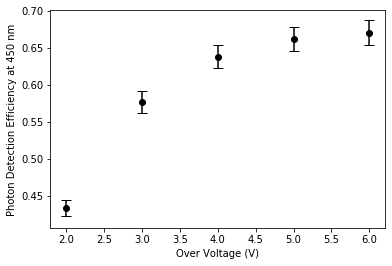}\par 
\end{multicols}
\end{center}
\caption{LEFT: PDE as a function of wavelength at a 3 V over voltage and 365 nm,  400 nm,  450 nm,  525 nm,  560 nm, 660 nm, 720 nm, 810 nm, and 900 nm. RIGHT: PDE as a function of over voltage at 450 nm.}
\label{fig:PDE}
\end{figure}

We measured PDE as a function of wavelength and over voltage using the charge integration method used previously to characterize the S13360-3050CS MPPC for the UFA telescope\cite{Li_2019b}. The PDE as a function of wavelength at a 3 V over voltage and the PDE as a function of over voltage at 450 nm can be seen in Fig. \ref{fig:PDE}. We find a PDE of (55 $\pm$ 3) $\%$ at 450 nm and a 3 V over voltage. 

\subsection{Linearity and Saturation}

At low irradiances, a SiPM will output a count rate directly proportional to incident photon rate. However, at high irradiances, incident photons begin to arrive with time separations shorter than the recovery time of the SiPM and cause the output of the SiPM to deviate from linearity. As many SiPMs are designed for low light environments but our application involves a high sky background, identifying the linear range and saturation of the S14160-3050HS MPPC is an important check to determine its feasibility for astrophysical observations. 

\begin{figure} [H]
\begin{center}
\begin{multicols}{2}
\includegraphics[height=5.5cm]{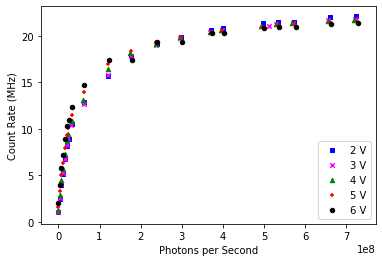}
\includegraphics[height=5.5cm]{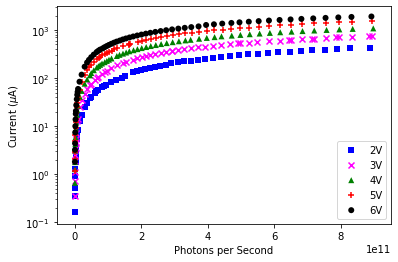}\par 
\end{multicols}
\end{center}
\caption{LEFT: S14160-3050HS MPPC count rate as a function of incident photon flux and over voltage with the Citiroc 1A evaluation system. RIGHT: S14160-3050HS MPPC current as a function of incident photon flux and over voltage without the Citiroc 1A evaluation system.}
\label{fig:linearity}
\end{figure}

We illuminated the SiPM using a 660 nm LED with increasing photon fluxes up to 7.2x10\textsuperscript{8} photons per second. The SiPM count rate using a 0.5 photoelectron threshold as a function of photon flux with the Citiroc 1A evaluation system and the SiPM current as a function of photon flux without the Citiroc 1A evaluation system can be seen in Fig. \ref{fig:linearity}.

We calculated a linear least squares fit for the linear region consisting of the first three data points that lie below 7.2x10\textsuperscript{6} photons per second for measurements both with and without the Citiroc 1A evaluation system. We find that the output count rate of the SiPM and Citiroc 1A combination deviates 10$\%$ from this fit at 1.6x10\textsuperscript{7} photons per second at a 3 V over voltage. We observe the SiPM and Citiroc 1A combination to saturate at 2.0x10\textsuperscript{8} photons per second which agrees with the Citiroc 1A maximum photon counting rate of 20 MHz provided by Weeroc\cite{datasheetCitiroc}. The output count rate of the SiPM and Citiroc 1A combination is proportional to over voltage until saturation where it then crosses over and becomes inversely proportional. This crossover could be due to a PDE proportional to over voltage and the inability of the Citiroc 1A evaluation system to distinguish between overlapping pulses. Without the Citiroc 1A evaluation system, we find that the current of the SiPM deviates 10$\%$ from the linear least squares fit at 4.8x10\textsuperscript{10} photons per second at a 3 V over voltage. The SiPM saturates at approximately 1.0x10\textsuperscript{11} photons per second.

\section{Conclusion}

We refined the characterization procedure used to previously characterize the S13360-3050CS MPPC for the UFA telescope and obtained a 3 V over voltage baseline characterization for the S14160-3050HS MPPC. We observe that the S14160-3050HS MPPC saturates well above our expected sky background SiPM count rate of 180 kHz with and without the Citiroc 1A evaluation system and is expected to operate linearly during observations. At a 3 V over voltage, we find a significant trade off between a high dark count rate of 1.08 MHz and a high PDE of 55 $\%$ at 450 nm. While the dark count rate for the S14160-3050HS MPPC is more than twice the dark count rate of the S13360-3050CS MPPC, the PDE is 25 $\%$ higher at 400 nm\cite{Li_2019b}.  At room temperature, astrophysical observations using the S14160-3050HS MPPC would primarily be limited by its dark count rate. As cooling can reduce dark count rate, future work will involve adding a cooling system to our characterization test stand to measure the temperature dependency of dark count rate.

We measured crosstalk probabilities and their slope to be approximately twice and 18 $\%$ greater than those reported in the S14160-3050HS MPPC datasheet\cite{datasheet}, respectively. The crosstalk probability of the S14160-3050HS MPPC is also approximately twice that of the S13360-3050CS MPPC at a 3 V over voltage\cite{Li_2019b}. To check for light leaking into the dark box that could be potentially affecting our measurements, we retook the crosstalk probability measurements after 1) shielding the SiPM with another smaller box, 2) turning the lights in the experiment room on and off with and without the shielding, and 3) measuring the irradiance inside the dark box at the location of the SiPM with lights inside the experiment room turned on and off using the same photodiodes used to characterize the SiPM. We did not find a significant change in crosstalk probabilities or the irradiance inside the dark box with the experiment room lights on or off given the minimum 20 pW limit of the photodiode.

In addition, after the conclusion of this study a different SiPM of the same model was mounted on the NUTTelA-TAO and used to measure the sky background and various stars at the Assy-Turgen observatory\cite{Lau_2020_10.1117/1.JATIS.6.4.046002}. 
Taking the crosstalk probability slopes of 5 $\%$/V from the Hamamatsu datasheet\cite{datasheet} and 5.9 $\%$/V from this study, a 21 $\%$ crosstalk probability at a 3 V over voltage would translate to crosstalk probabilities at a 2.7 V over voltage of approximately 19.5 $\%$ and 19.2 $\%$, respectively. From Fig. 7 in Lau et al. 2020\cite{Lau_2020_10.1117/1.JATIS.6.4.046002}, both these crosstalk probabilities would correspond to a 0.5 photoelectron threshold count rate of approximately 1.1 MHz which agrees with the measurements in this study. We conclude that the high crosstalk probability found in this study could potentially be due to a higher than expected afterpulsing and delayed crosstalk effect or a difference between how crosstalk probability was defined and calculated in the datasheet and this study.



Based on the characterizations performed in this study, we find that the S14160-3050HS MPPC meets the required specifications for the UFA telescope. However, as the S14160-3050HS MPPC has a dark count rate at room temperature an order of magnitude higher than our anticipated sky background, the sensitivity of our observations would be dark count limited. If cooling and a coincidence scheme can significantly reduce the number of false alarms caused by dark counts and crosstalk, then the S14160-3050HS MPPC would be more advantageous for the UFA telescope than the S13360-3050CS MPPC. These results can be extended to future ground-based observatories with similar sky backgrounds and specifications in sub-second time domain astrophysics.

\acknowledgments 
 
This work was supported by the Hong Kong University of Science and Technology. S.L. would like to thank Albert Wai Kit Lau and Ulf Lampe for their hospitality and support with equipment.
\bibliography{report} 
\bibliographystyle{spiebib} 

\end{document}